# The emergence of sequence-dependent structural motifs in stretched, torsionally constrained DNA

Jack W Shepherd[1, †], R J Greenall[1], M I J Probert[1], Agnes Noy[1*], Mark C. Leake[1,2*]

[1] Department of Physics, University of York, York, YO10 5DD, UK.

[2] Department of Biology, University of York, York, YO10 5NG, UK.

* To whom correspondence should be addressed, Email: mark.leake@york.ac.uk
Tel: +44 (0)1904 322697; Fax: +44 (0)1904 322214.
Correspondence may also be addressed to Agnes Noy, Email: agnes.noy@york.ac.uk

[†] Present Address: Jack W Shepherd, Cell Polarity Migration and Cancer Unit, Institut Pasteur, 25-28 Rue de Dr Roux, 75015 Paris, France.

**ABSTRACT**

The double-helical structure of DNA results from canonical base pairing and stacking interactions. However, variations from steady-state conformations result from mechanical perturbations in cells. These different topologies have physiological relevance but their dependence on sequence remains unclear. Here, we use molecular dynamics simulations to show that sequence differences result in markedly different structural motifs upon physiological twisting and stretching. We simulated overextension on four different sequences of DNA ($(AA)_{12}$, $(AT)_{12}$, $(GG)_{12}$ and $(GC)_{12}$) with supercoiling densities within the physiological range. We found that DNA denatures in the majority of stretching simulations, surprisingly including those with overtwisted DNA. GC-rich sequences were observed to be more stable than AT-rich, with the specific response dependent on base pair ordering. Furthermore, we found that $(AT)_{12}$ forms stable periodic structures with non-canonical hydrogen bonds in some regions and non-canonical stacking in others, whereas $(GC)_{12}$ forms a stacking motif of four base pairs independent of supercoiling density. Our results demonstrate that 20-30% DNA extension is sufficient for breaking B-DNA around and significantly above cellular supercoiling, and that the DNA sequence is crucial for understanding structural changes under mechanical stress. Our findings have important implications for the activities of protein machinery interacting with DNA in all cells.

**INTRODUCTION**

In the cell, DNA is constantly under mechanical perturbation from a range of proteins as well as protein/nucleic acid based molecular machines. These perturbations are necessary for a wide range of DNA functions, including replication, repair, gene expression, and chromosomal packaging[1], [2]. Single-molecule force spectroscopy experiments have shown that DNA can be overstretched from 10% up to 70% beyond its relaxed contour length at a nearly constant force of around 65-70 pN [3], [4]. This force-derived plateau has been associated with a structural transition from the canonical B-DNA to the extended form called S-DNA [2] that has been observed to be dependent on a number of different factors. It was observed that AT-rich DNA, in particular poly-d(AT) fragments, favoured the extension to the S-form in contrast to GC-rich or poly-d(CG) [5]. In terms of torsional stress, unwinding greatly reduces the force at which the B–S transition occurs and overwinding makes it harder to happen [6]. For significantly overtwisted DNA (with a supercoiled density $\sigma > 0.037$), stretching DNA with magnetic tweezers showed the emergence of a Pauling-like DNA (P-DNA) structure, which was deduced to have the bases flipped outside and the two helical backbones grouped inside [7].

However, it is difficult to visualize experimentally the effect of stretching and twisting DNA at the base pair level, for example using the most advanced dynamic super-resolution fluorescence microscopy methods for DNA whose spatial localisation precision for pinpointing individual reporter dye molecules is at best a few tens of nm [8]–[10]. This level of precision is equivalent to the separation of ~100 base pairs (bp) and so struggles to generate information about the orientation or specific interactions of any given nucleotide. Structural transitions have been discerned through force and torque considerations but the structures themselves are not directly visible [11]. By contrast, computer simulation works at its best over short length scales, being able to provide high-resolution atomic level precise descriptions of the different overstretched DNA states, allowing identification of specific structural motifs which may emerge. These detailed descriptions can be correlated with experimental data, often at higher length scales, to build a holistic understanding of the structural dynamics of DNA molecules.

Atomistic molecular dynamics (MD) initially described S-DNA as a ladder-like structure where Watson-Crick (WC) hydrogen bonds were mostly maintained and the base-base intra-strand stacking was preserved through a high inclination of the base-pairs with respect to the molecular axis or substituted by stacking between bases from opposite strands (i.e. inter-strand) [12]–[14]. Later, it was demonstrated that S-DNA consists of a full umbrella of conformations where the S-ladder motif can coexist with short melting bubbles generated especially on AT-rich segments [15]–[21]. Similarly, computer simulations described in atomic detail (and confirmed) the structural appearance of the overstretched and overtwisted form of P-DNA [22].

The effect of torque on simulated linear DNA structures has been relatively underexplored in comparison. By imposing torsional constraints, simulations have shown that unwounded DNA has a greater predisposition to form melting bubbles compared to relaxed DNA, with this propensity being strongly dependent on sequence: denaturation was only observed on an (AT)$_3$ segment in linear DNA [23] and prominently in AT-rich areas in undertwisted small DNA circles between 60-110 bp [24]. Modelling an infinite linear tract of DNA upon undertwisting showed that DNA partitions its perturbations such that there are regions of extreme disruption in coexistence with regions which remain relatively close to the canonical B-form DNA structure, and that this is sequence-dependent [25].

Atomistic MD is not the only means of simulating DNA. Although atomic simulations have the capacity for generating highly resolved structural details, they are computationally very demanding and so the extent of time sampling is usually limited. Coarse-grained models like oxDNA and 3SPN [26], [27], which describe a DNA molecule with significantly reduced degrees of freedom, are an alternative, powerful tool for describing the mechanism of overstretching and longer length and time scales with the caveat of sacrificing some level of spatial precision [28], [29]. Force-induced melting was initially described by oxDNA [28], and later the combination between denaturation bubbles and S-DNA was observed using a 3SPN-based potential [29].

However, atomistic MD is a very convenient means of interrogating precise base pair level structural motifs. Despite this, the combined effects of precisely imposed stretching and twisting perturbations together with the dependence on different DNA sequences has not been studied in depth to date. In our present work, we report a comprehensive suite of atomically precise molecular dynamics simulations in order to interrogate the structural impact of DNA overextension on topologically-constrained molecules within the range of normal supercoiling density observed *in vivo* (σ±0.068) [30], [31]. We used four different sequences (poly-d(A), poly-d(AT), poly-d(C), poly-d(CG)) which cover a significant variety of known DNA flexibility [32]. Additionally, these sequences cover two binding motifs known to be biologically relevant: the TATA box [33] which is used for transcription factor binding, and the CGCG box which is known to be part of many signalling pathways in plants [34] and which can occur within CpG islands, regions of high CG density which are found in approximately 40% of mammalian gene promoter regions [35].

We focused on the early-stage overstretching regime (just up to the level of 40% beyond the relaxed contour length) as this is the upper known limit of mechanical distortion that can be achieved through protein-binding (for example, in the TATA-binding protein [36], and the recombinase enzyme RecA [37]) and thus enabled us to explore the most physiological regime. We found that this level of overstretching was sufficient to disrupt the B-DNA form in spite of the different torsional constraints, and that melting bubble behaviour and non-canonical structure formation were found to be significantly dependent on sequence.

## MATERIAL AND METHODS
### Software simulation platforms
All simulations were set up with the Amber 17 suite of programs and performed using the CUDA implementation of Amber's pmemd program [38]. The initial 24 bp structures for four different linear DNA fragments were obtained using Amber's NAB utility [38] for the following sequences: $(AA)_{12}$, $(AT)_{12}$, $(GG)_{12}$ and $(GC)_{12}$ for exploring the effects of sequence dependence. Note that sequences do not have GC caps at each end because terminal base pairs are under constraints in all simulations so cannot suffer from helix-end melting. The molecular forcefield used for the DNA was the Amber Parm99 forcefield [39] with bsc0 [40] and bsc1 [41] corrections for backbone dihedrals.

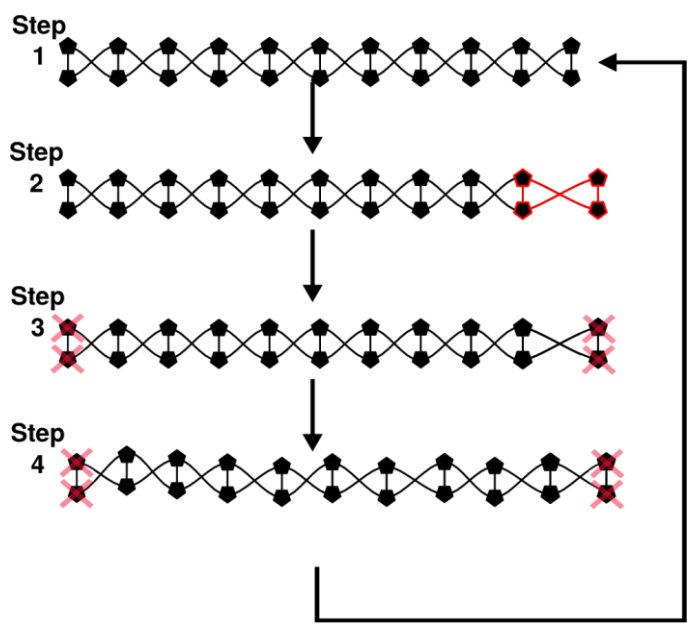

**Figure 1:** Schematic of stretching methodology. Step 1: an initial structure is obtained. Step 2: a terminal base pair (highlighted red) is displaced from its starting position by 1 Å. Step 3; the structure is minimized with harmonic constraints at both ends (pink crosses). Step 4: the system is subjected to a production run of 0.5 ns and the final frame is used as the input for the next round of stretching, minimizing and modelling.

**Simulations in implicit solvent**

The Generalized Born (GB) model [42], [43] was applied to define the solvent implicitly, together with an effectively infinite long-range electrostatic cut-off, the latest GBneck2 corrections and mbondi3 Born radii set for a better reproduction of molecular surfaces, salt bridges and solvation forces [44]. Systems were minimized using a combination of steepest descent and conjugate gradient methods. Simulations were run at constant temperature (300 K), maintained by the Langevin thermostat [45], using an effective salt concentration of 50 mM determined by the Debye−Huckel screening parameter. Integration time steps were set at 1 fs with the DNA structure written to disk every 1 ps.

To stretch DNA while maintaining torsional constraints, fixed harmonic traps with an effectively infinite associated constant of 500 kCal/Mol/Å$^2$ were applied to all atoms in the terminal base pairs (see Figure 1). DNA was stretched on a series of umbrella sampling trajectories, where one of the terminal base pairs was moved 1 Å from the centre of the molecule along the helical axis (Figure 1). The final frame of the previous umbrella sampling window was used as the starting structure for the next stretching event (Figure 1). Through analysis of temperatures and potential energies, the system was found to be almost instantly equilibrated after the perturbation (see Figure S1) as the increment on each extension step falls within the range of end-to-end thermal fluctuations found in a piece of unconstrained DNA [46]. Every simulation window was 500 ps in width, using an overall stretch rate of 2 Å/ns. On a single umbrella sampling trajectory, each DNA molecule was extended by a total of 30 Å, resulting in an increase in contour length of approximately 39.6% and a total run time of 15 ns. Each of the four DNA sequences considered was modelled using nine different supercoiling densities spanning a range used in previous magnetic tweezers studies (σ values of: 0, ±0.017, ±0.034, ±0.051, and ±0.068), whose negative extremity is also comparable to that estimated for live bacteria [31]. This led to 540 ns of simulation time in total.

**Trajectory analysis**

Trajectories were analysed using cpptraj [47] to find canonical and non-canonical hydrogen bonds using distance and angle cut-offs of 3.5 Angstroms and 120° respectively. The esander routine was used to extract Van der Waals interactions between successive nucleotides to find canonical stacking interactions, and between a nucleotide and every non-neighbour nucleotide for the non-canonical stacking energies. Energy calculations here were all performed in implicit solvation with the same forcefield and salt conditions as were used during the simulation. For each base pair, hydrogen bonds and stacking energies were averaged for each stretch extent in each simulation and these values were used to describe the structural details of DNA stretching and to quantify the presence of melting bubbles (defined to be two or more sequential base pairs which have on average less than one canonical WC hydrogen bond present). Representative structures were then extracted using VMD [48] or Chimera [49]. Two base pairs at each end of the fragment were not considered in the trajectory analysis to avoid end-effects.

**Simulations in explicit solvent**

For ensuring accuracy, key structures from implicit solvent simulations were simulated for 5 ns in explicit solvent. The final frames of the following umbrella sampling windows were chosen as initial structures for the subsequent simulations: $(AA)_{12}$ with a 2 nm stretch, $(AT)_{12}$, $(GG)_{12}$, and $(GC)_{12}$ with a 2.5 nm stretch, all on σ = ±0.068. These initial structures were solvated by a box of TIP3P water [50], neutralized and surrounded by extra 200 mM NaCl [51], [52] Before production simulations the solvated and relaxed systems were equilibrated using a standard protocol [53]. Simulations were carried out with an electrostatic cut-off of 1.2 nm, in the NPT ensemble using the Langevin thermostat and the Berendsen barostat [54] and using a 0.5 fs integration time. The full 5 ns simulations were analysed for hydrogen bonding and VdW stacking interactions to assess conformational drift away from the implicit solvent simulation, and the final 1 ns of the simulation was used to create an average structure. Trajectory analysis was performed as for the implicit solvent simulations using the GB solvation model with the Debye-Huckel parameter set to simulate 200 mM salt concentration.

# RESULTS

## Sequence and stretch dominate over twist for the emergence of melting bubbles

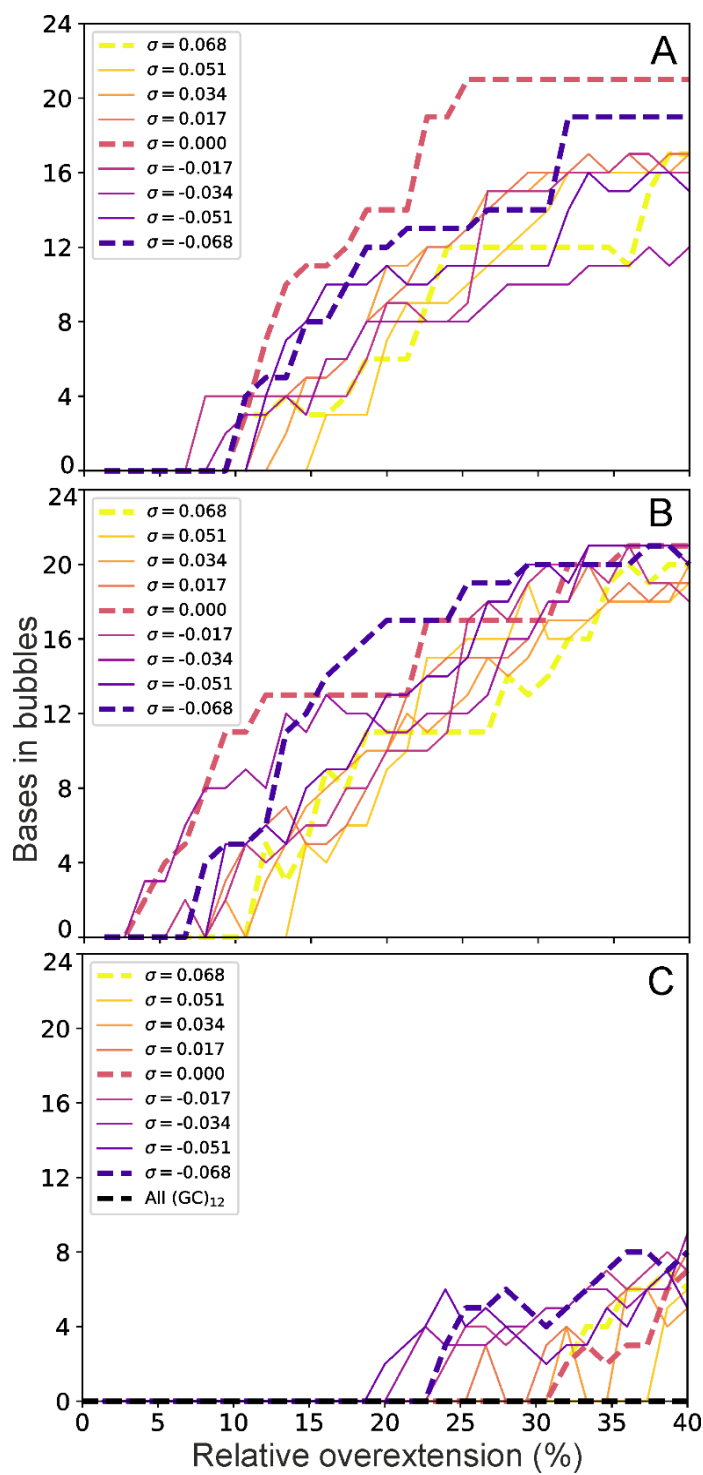

For gaining a general perspective of the degree of disruption of the B-DNA state among the whole set of simulations, we quantified the emergence of denaturation events. Figure 2 shows the effect of supercoiling and sequence on melting bubble formation through the quantification of the number of bp that have lost their canonical hydrogen-bond interactions for a particular stretching step (see Materials and Methods). Both AT-rich sequences present melting bubbles early in the stretching process (before 20%) that end up disrupting more than half of the DNA construct in the maximum extension. This behaviour was found to be generally shared on all supercoiling densities, suggesting a small dependence on torsional stress and a high level of fragility under stretching. In contrast, poly-d(C).poly-d(G) is the more resistant sequence as it does not form melting bubbles for any supercoiling density. The strength of stacking interactions [55], [56] and three hydrogen bonds between C and G are the probable reasons for preventing considerable perturbations. Poly-d(CG).poly-d(CG) presents a significantly different behaviour compared to poly-d(C).poly-d(G) as it forms melting bubbles on all the simulations affecting a minimum of four bp. It is also the sequence with the most supercoiling dependence: the earliest onset of a melting bubble is at an extension of ~20% and occurs for σ=-0.051, while the last supercoiling density to create a melting bubble is σ=0.051 at an extension of ~38%, meaning that a changed supercoiling density allows the DNA to absorb twice as much extension before melting. This is in line with what could be expected – negative supercoiling tends to open the helix up and make it easier for base pairs to dissociate, whereas positive supercoiling is associated with packing the base pairs more tightly and constraining them in place. Significantly, there is only ever one melting bubble produced in the (GC)$_{12}$ fragment, demonstrating that the perturbation is efficiently localized and grows in place rather than producing new bubbles (see further Results section).

**Figure 2.** Number of bp involved in a melting bubble for each of the molecular dynamics simulations performed. A) $(AA)_{12}$ presents melting bubbles for undertwisted structures as early as 7% overstretching, while overtwisted constructs form bubbles prior to 15% extension. B) $(AT)_{12}$ denatures in places almost immediately, for undertwisted structures at 2% overextension in the earliest case, while the last structure to form a melting bubble is overtwisted by σ=0.051 and maintains its B-DNA structure until around 13% extension. C) $(GC)_{12}$ is considerably more stable with undertwisted structures forming melting bubbles only after 20% extension and DNA with σ=0.051 maintaining B-DNA hydrogen bonding up until 37% overstretching. Also C) $(CC)_{12}$ is stable in the B-DNA form for all extensions and all σ and is therefore the most stable structure. Note that the minimum of bp involved in a single bubble are two and that the total number of bp can be originated from one long melted DNA stretch or several small denaturation events. For each sequence, the bubble size for σ=-0.068, σ=0, and σ=0.068 are plotted with dashed lines.

However, in these physiologically relevant supercoiling regimes even positively supercoiled DNA can form melting bubbles in a sequence- and extension-dependent manner. Thus, in general terms, our simulations show that the formation of denaturation bubbles is dependent on applied torsion only in a moderate way, suggesting that at these extensions the overstretching strain dominated over the stress due to twist.

In terms of sequence content, these results confirm the long-standing view that AT-rich sequences are more fragile compared with GC-rich sequences, although they also point to a far more complex picture than had previously been envisioned due primarily to the importance of stacking forces. Indeed, beyond melting bubble formation, the generation of stable structural motifs is entirely sequence specific (see the following Results sections), with the behaviour of poly-d(A) and poly-d(C) strikingly different from that of poly-d(AT) and poly-d(CG), respectively, indicating that simple approximations of DNA behaviour based on AT/GC content are reliable for comparatively long DNA sequences only in situations such as DNA melting for PCR considerations and single-molecule buckling point determinations [57] analysis.

**The poly-d(A).poly-d(T) sequence generates flexible dual motifs which may function as shock absorbers**

Figure 3 shows the evolution of non-bonded interactions as the poly-d(A) sequence is stretched at the negative supercoiled density of -0.068. We found that canonical hydrogen bonds started to be significantly disrupted around a relative overextension of ~10%, with completely disrupted base pairs seen at 15% extension. Simultaneously, the disrupted WC hydrogen bonds begin to be substituted by non-canonical hydrogen bonds. At around an extension of 35%, non-canonical hydrogen bonds largely disappear from the central region of the DNA indicating a far more melted configuration, though they persist at one end, likely due to edge effects from the immobilized base pair. As the WC hydrogen bonds are lost and non-canonical ones formed, stacking interactions follow suit. Canonical stacking energies reduce while non-canonical stacking increases, as may be expected for a DNA molecule whose bases are hydrogen bonding with unusual partners.

The structure extracted at the overstretch step of 24% (Figure 3) shows detail of the denatured conformation, indicating two separate but linked modes of structural adaptation to the applied mechanical perturbation. On one hand, the central region of the DNA fragment (between bases 11 and 14) contains a low number of non-canonical hydrogen bonds and stable non-canonical stacking interactions. This low hydrogen bonding area is highlighted in blue on Figure 2 and presents a region with a collapsed backbone and associated rotation of the bases themselves stacking/hydrogen bond interplay does not persist beyond an extension of 35%. This is well expected for the sequence with some of the weakest stacking interactions (see for example [55], [56]) as well as fewest number of WC hydrogen bonds: the structure simply does not have sufficient stability to form a new, lower-energy structure, and thus melts. For the trajectory containing this sequence at the highest level of overtwisting (σ=+0.068, see Supplementary Figure S2) we also observe denaturation bubbles although to a lesser extent. The structure is more stable in the early stages of overstretching as is expected from previous work[58]. However, there are disruptions to the WC hydrogen bonding at ~10% strain, and at ~20% strain an isolated melting bubble briefly comes into existence. It lasts only a few more stretching events, however, before being stabilized. Finally at 25% strain the central region of the DNA is significantly disrupted, and at this point the melting bubble persists and grows.

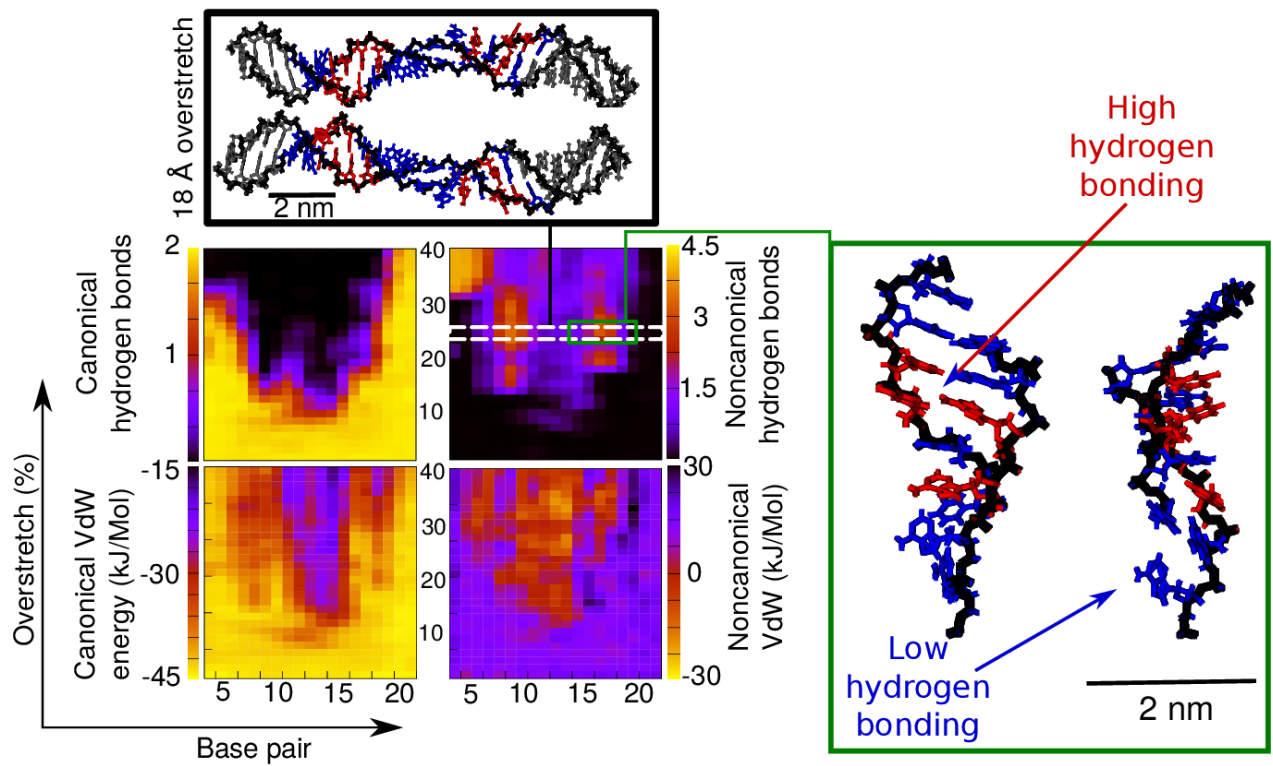

**Figure 3:** Canonical/non-canonical hydrogen bonds and stacking energies (see methods) as a function of applied overstretch (up to 40%) on the simulation of poly-d(A) at σ=-0.068. The insets show representative structures for an overstretch of approximately 24% where blue indicates a bp with 1-2 non-canonical hydrogen bonds and red indicates a bp with >2. The two images shown inside the black and green boxes are two side views of the same structure rotated 90 degrees.

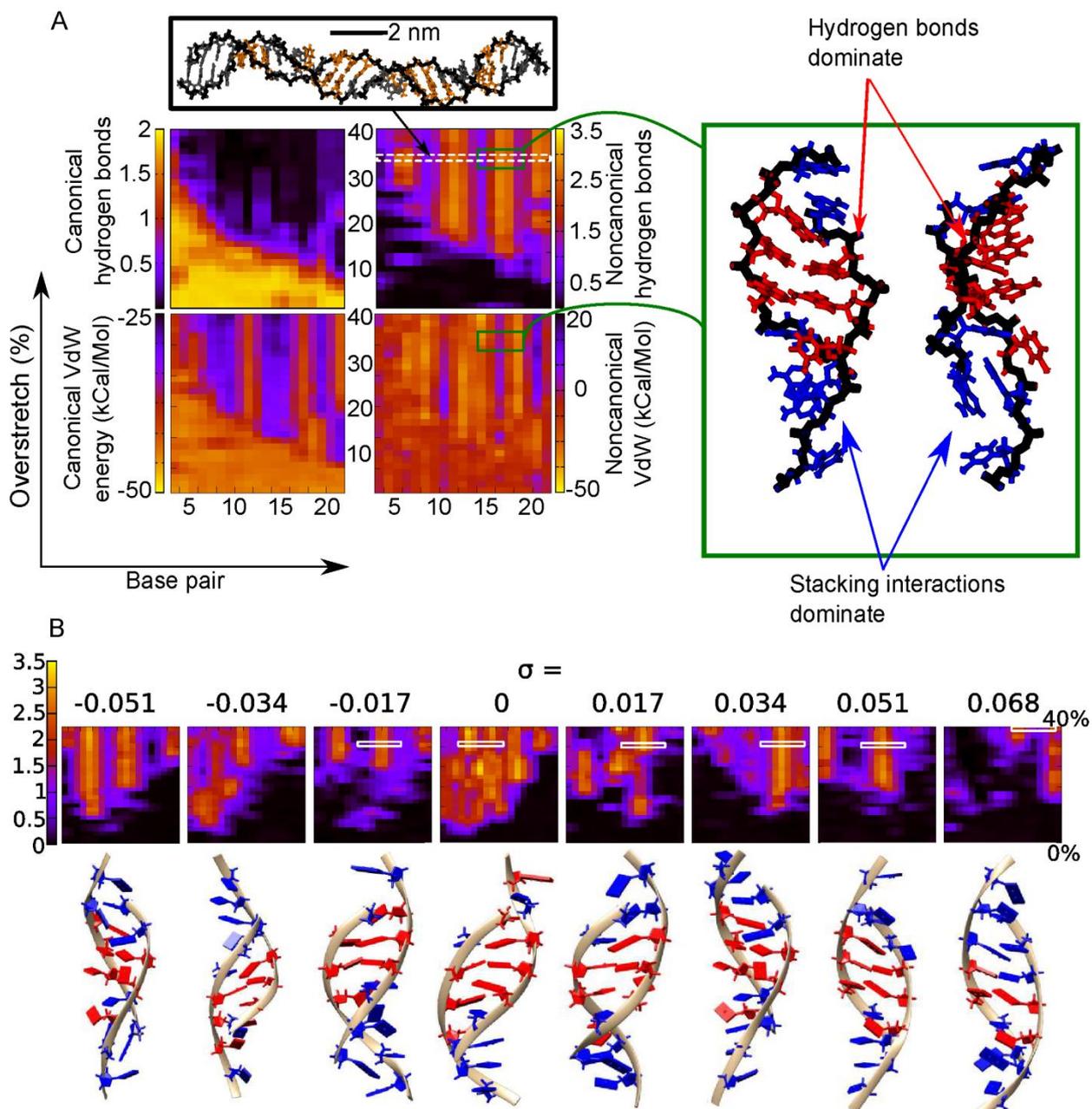

**Figure 4: A)** Canonical/non-canonical hydrogen bonds and stacking energies (see methods) as a function of applied overstretch (up to 40%) on the simulation of poly-d(AT).poly-d(AT) at σ=-0.068. Insets show representative structures for an overstretch around 32%. The two images shown inside the green box are two side views of the same structure rotated 90 degrees. Here blue indicates areas with <2 non-canonical hydrogen bonds and <-25 kCal/Mol non-canonical stacking. The red indicates >2 non-canonical hydrogen bonds and >-25 kCal/Mol non-canonical stacking. **B)** Emergence of non-canonical hydrogen bonds as a function of applied stretch and supercoiling density together with the corresponding structural motifs on the sequence poly(AT).

Non-canonical hydrogen bonds are not seen in the same areas, and the inset structure in Supplementary Figure S2 shows a significantly disordered structure coexisting with B-DNA like regions. Meanwhile, the canonical stacking energies are significantly reduced, while the non-canonical energies increase in magnitude. Similarly to the undertwisted case in Figure 2 the DNA reduces its total energy by engaging in new non-bonded interactions. With both the undertwisted and overtwisted structures forming collapsed-backbone structures, it seems likely that with increased extension a qualitatively P-DNA like state with exposed bases could be reached, albeit with a significantly greater helical pitch, and suggests an alternative pathway for exposing and manipulating short sequences of bases.

**The poly-d(AT).poly-d(AT) sequence generates stretch-resistant motifs stabilized by stacking and non-canonical hydrogen bonds**

The poly-d(AT)·poly-d(AT) sequence showed a similar tendency to melt as the poly-d(A)·poly-d(T) under supercoiling density σ = -0.068 (see Figure 2), but the non-canonical interactions we found to be significantly more stable with applied extension (see Figure 4). As in the poly-d(A) fragment, non-canonical hydrogen bonds appear where canonical ones have broken, replacing the WC interactions entirely, and the stacking forces follow the same trend. At an extension of 17%, the DNA molecule forms a structure with a regular pattern of increasing and decreasing non-canonical hydrogen bonds, suggesting a new stable structural motif. Two distinct states are clearly seen in the structure. The first type of regions is 6 nucleotides long, it is stabilised by non-canonical hydrogen bonds (highlighted in red on Figure 4A) and it has a B-DNA like macrostructure although the bases are slightly flipped away from the helical axis. The second type of structural motif is shorter (~3 base pairs) and is dominated by stacking interactions, presenting low hydrogen bonding (highlighted in blue on Figure 4A). The bases on opposite strands tend to be interlinked, suggestive of a conformation similar to zip-DNA [59]. Our results suggest that this dual coexisting method of duplex stabilization is mechanically stronger than the previous structural motif emergent in poly-d(A) as it persisted until the end of the simulations. For the overtwisted simulations (see Supplementary Figure S3), the picture is less clear. The WC base pairing is lost early in the simulation, though later than in the undertwisted case (Figures 2 and 4). A significant central region of the construct maintains the WC pairing and canonical stacking of its B-DNA structure, but this is flanked by regions of increased non-canonical VdW energies and non-canonical hydrogen bonding. In the inset in Supplementary Figure S3, the dual stabilization modes as in Figure 4 can be clearly seen though they are established considerably later, at ~35% extension. Unlike poly-d(A), there is a melting bubble initiated at around 20% strain which persists until the end of the simulations (see left hand inset in Supplementary Figure S2) in which there is stabilization only through the non-canonical hydrogen bonds. This shows that the stabilization modes seen always to coexist in Figure 4A can be generated singly also. Whether this is supercoiling-dependent is unclear, but it seems likely that it plays a role given that the motifs appear always in tandem in undertwisted fragments while the only single motif occurred in the overtwisted DNA (see Figure 4B).

Once again, in the overtwisted simulation, there are individual bases seen flipped outwards as a consequence of a collapsing backbone, once again demonstrating that a pathway to a bases-exposed conformation other than significant positive supercoiling exists early in the overstretching regime in AT rich DNA.

**The poly-d(C).poly-d(G) sequence has a high resistance to mechanical perturbation due to strong canonical stacking and hydrogen bonding**

The poly-d(C)·poly-d(G) is the most stable of all the sequences that we simulated here. All stretching simulations constrained at different σ maintain most of the canonical non-bonded interactions as seen in Figure 5 and Supplementary Figure S4. Base pairs adopt a steeply inclined configuration in a similar way as was previously described for S-DNA [2], [8], [10]. We observed an increase in the non-canonical stacking interactions as the DNA is being pulled due to the more prominent role of the diagonal interactions (or cross-stacking), although these changes are much less dramatic and cannot be related to any other configuration apart from S-DNA. Similar S-DNA like configurations were observed on over-winding DNA (see Supplementary Figure S4 and inset structures), though there is a smaller level of structural alteration as opposed to under-twisted DNA. This indicates the tighter packing of highly twisted bases is a substantial factor in regards to determining the level of structural stability. In summary, it seems here that the canonical stacking and the three WC hydrogen bonds are largely sufficient to stabilize this level of perturbation.

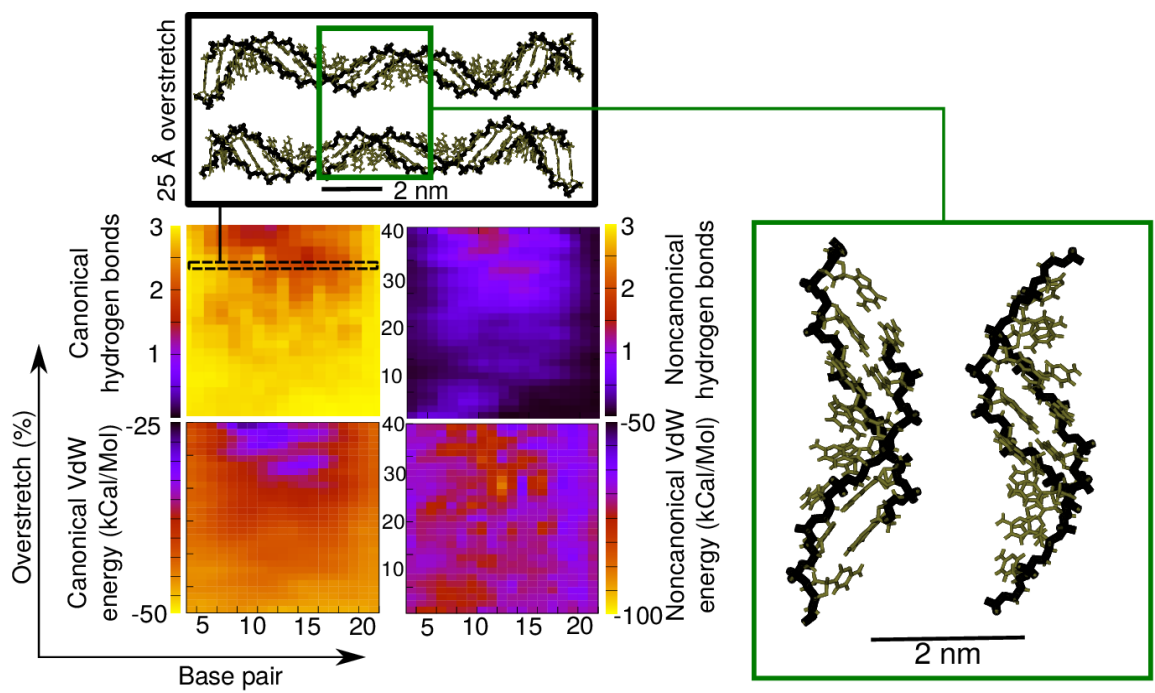

**Figure 5:** Canonical/non-canonical hydrogen bonds and stacking energies (see methods) as a function of applied overstretch (up to 40%) on the simulation of poly-d(C).poly-d(G) at σ=-0.068. The insets show representative structures for an overstretch of 2.

**The poly-d(CG).poly-d(CG) construct forms mechanically stable four bp motifs which may accommodate significant overextension**

We found that the behaviour of the poly-d(CG)·poly-d(CG) sequence was significantly different from that observed for the poly-d(C)·poly-d(G) sequence, in spite of the CG content being the same. The principal difference here is the presence of intra-strand stacking of RY and YR bases instead of YY and RR (Figure 6). At an extension of 14% we observed backbone collapse similar to that of the AT-rich sequence, with a resultant loss of around two of the canonical hydrogen bonds in favour of non-canonical bonds. As seen for other sequences, there is a corresponding increase in non-canonical stacking energy as canonical hydrogen bonds are lost. The structural details of Figure 5 indicate that some of the two opposite bases are interacting via inter-strand stacking instead of the WC hydrogen bonds.

This interaction seems to facilitate the formation of a four bp motif, which consist of two bases on one strand followed by two bases from the opposite strand and is similar to what have been described in previous studies [15], [36]. This quadrubase configuration is characterized by a high inclination of the bases almost perpendicular to their usual orientation and maintains the stacking order CGCG, which is maximally strong [55], [56]. The disruption here preserves the canonical structure in other parts of the fragment, and the novel conformation appears able to absorb significant additional perturbation, as it does not grow in size through the simulation.

For the overtwisted case (Supplementary Figure S5), the behaviour is qualitatively similar to that indicated in Figure 6 though with σ = 0.068 the structure is considerably more resilient to applied strain, with WC base pairing being lost and non-canonical interactions appearing after around 30% extension. As in Figure 6, the loss of WC hydrogen bonds and canonical VdW stacking correlates exactly with the rise in non-canonical hydrogen bonding and stacking. Once again, indicated in the inset in Supplementary Figure S5, the 4 base pair motif reminiscent of [15], [36] clearly develops and is stable to the end of the simulation.

That the 4 base pair motif is strongly conserved is an important, unexpected result. While in other sequences the produced motifs were similar (see Figure 3 and Supplementary Figure S2, and Figure 4 and Supplementary Figure S3) in this case the motif produced is identical in every respect. Four bases tilt such that they are almost aligned with the helical axis, and stack around the outside of a collapsed backbone. These motifs appear separate and stable on their own. Perfect replication of this motif in the over and undertwisted DNA suggests that there could be an important a biological role either in stabilizing the DNA or in exposing the base sequence for replication, regulation, or repair, especially in mammalian promoter sites with significant CpG islands [35] and plant cell signalling pathways making use of a CGCG box [34] The motif we observe could be a mechanism for recognition regulation in cell in either of these cases.

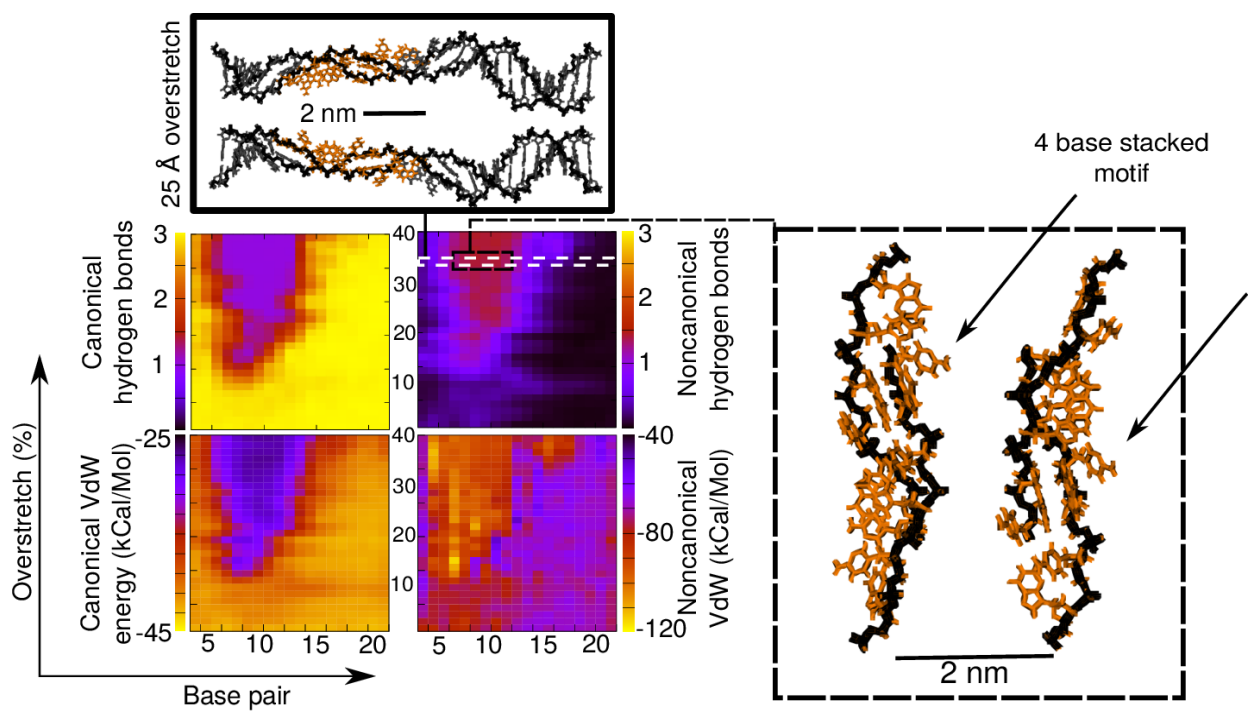

**Figure 6:** Canonical/non-canonical hydrogen bonds and stacking energies (see Materials and Methods) as a function of applied overstretch (up to 40%) on the simulation of poly-d(CG).poly-d(CG) at σ=-0.068. The insets show representative structures for an overstretch of 25 Å (approximately 32%) and the two image inside each box are two side views rotated 90 degrees.

**Simulations in explicit solvent**

In order to demonstrate that the structures extracted from the full-length implicit solvent routines were not an artefact associated with the approximation from the solvation model, 5 ns of explicitly solvated molecular dynamics was performed (see Materials and Methods), using as a starting structure the final frame of the appropriate implicit solvent umbrella sampling window. The structures chosen were $(AA)_{12}$ with extension 25%, $(AT)_{12}$ with extension 32% and $(GC)12$ with extension 32% as in these cases unexpected behaviour was observed in the implicit solvent simulations. In each case the two maximally torsionally stressed states were simulated (i.e. σ = ±0.068).

Supplementary Figure S7 shows the hydrogen bonding and stacking interactions for the 5 ns simulation of each structure. In general, the drift from the starting structure is very low indicating that the structures found in the implicit solvent simulation are a stable minimum energy state. The key features from the implicit solvent simulations of all structures remain in place. Specifically, the overtwisted AT rich structures continue to show melting bubbles, and the $(AA)_{12}$ simulation maintains the dual stacking/hydrogen bonding states. $(AT)_{12}$ retains the distinctive alternating motifs, indicating that the partitioning of perturbation between stacking and hydrogen bonding is energetically favourable in physiological conditions and is not an artefact of the solvation system used. As in the implicit case, the $(GC)_{12}$ demonstrates a distinct edge between highly hydrogen-bonded and less bonded regions, and indeed between regions of high and low stacking energy.

The remarkable correspondence between the implicitly and explicitly solvated systems is indicative both that the implicit solvation simulations are reliable but also that the transitions seen are available in the living cell, at normal physiological stresses and ionic strengths, and therefore these structures may play a role an important role in the expression of genes.

**DISCUSSION**

In this study we have run an extensive set of molecular dynamics simulations deforming DNA within the biological regime of stretching and torsional forces for four different sequences in order to investigate the structural effects of the different type of perturbations and their sequence-dependence.

Our simulations focus on the early-stage of the overstretching reaction (just up 40% of extension), comparable to the range of DNA extension resulting from the activities of DNA-binding proteins, such as TBP and RecA [37]. Likewise, torsional constraints correspond to supercoiled densities between -0.068 and 0.068, which are the values observed in vivo either for prokaryotes [31] or eukaryotes [30]. In general terms, we observe that this level of mechanical stress disrupts the canonical B-DNA form in most of the cases. Numerous denaturation events occur, even for the most positively supercoiled DNA, suggesting a weaker influence of torsional stress compared with stretching perturbation, which seems to dominate in this physiologically relevant regime. The observed bases flipped away on overtwisted DNA could be the first step for the formation of a P-DNA like structure.

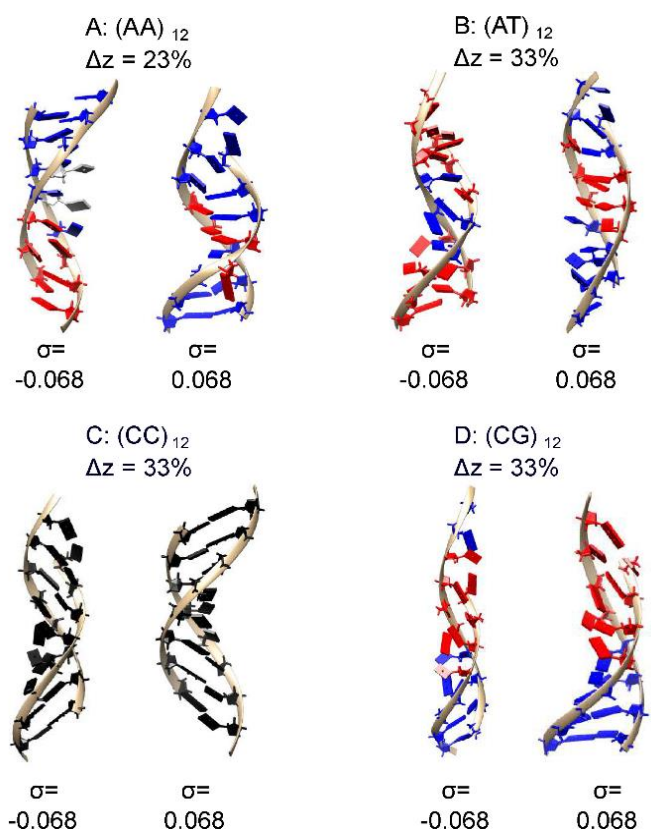

**Figure 7:** Structural motifs seen in the four DNA sequences with σ=±±0.068 extracted from different overstretching extensions. Blue indicates low non-canonical hydrogen bonding while red indicates high non-canonical hydrogen bonding; the distinct interfaces between the two regions are therefore visible. In C) black is used because the structure is essentially intact. Colouring is taken from the appropriate Figures in the main text and Figures S2-S5.

The four DNA sequences chosen for this study ($(AA)_{12}$, $(AT)_{12}$, $(GG)_{12}$ and $(CG)_{12}$) enable us to interrogate not only the impact of the overall GC/AT ratio but also the impact of the bp-order, which causes different strengths on stacking interactions, being these critical for DNA stability. Overall, AT-rich sequences show higher propensity to melt compared with CG-rich sequences in agreement with previous studies [15], [23], [24], [36], although striking differences are found whether sequences contain RY and YR intra-strand stacking or YY and RR.

Poly-d(C) appears to be significantly more resistant than the other G-rich sequence, poly-d(CG), because the former is the sole sequence that do not form melting bubbles for any supercoiling density, and the latter presents denaturation for all torsional constraints. An important point to note is that poly-d(CG) is the sequence that shows the highest sensitivity to changes of DNA twist, according to the wide range of observed extensions at which DNA starts to melt, being the first one at 20% for σ = - 0.051 and the latest at almost 40% for σ = 0.051. In contrast, for the two sequences with 100% AT content, the structure of the double helix is broken for all supercoiling densities at extensions as low as 15%, although the stability of the resultant conformation is notably different.

Figure 7 summarizes the key structural motifs emerged as a result of the applied mechanical stress. At first glance, we can differentiate between GC rich sequences, which develop a ladder-like geometry that can be classified under the umbrella of S-DNA, and AT rich sequences, that present more apparent bubbles with bases flipped away. Nevertheless, DNA molecules with an alternating RY and YR sequence pattern produce alternative stable structural motifs due to the higher stability of non-canonical stacking interactions. Since both RY/YR sequences studied here have biological importance in promotion and regulation of genes (the TATA box [36], CGCG box [34], and CpG islands [35]) it is likely that the specific motifs we found have biological roles. Specifically, both the TATA and CGCG boxes require recognition from an associated transcription factor, while the CpG islands are found in around 40% of promoter and exonic regions of mammalian genes. Mechanically stable structural motifs give a potential mechanism by which these regions may be activated or deactivated without significantly disrupting the local genome.

$(AT)_{12}$ and $(AA)_{12}$ polymers distort the helicoidal DNA shape with two alternating sequence motifs. The first one maintains a certain appearance to B-DNA although bases are slightly opened towards the major groove and stabilized by non-canonical hydrogen bonds. The second one makes a clear melting bubble by barely having hydrogen bonds, although we did observe some non-canonical stacking interactions. The different degree of strength in stacking forces seems to be the cause of poly-d(AT) higher resilience compared to poly-d(A).

A strong inclination of bp in respect to the molecular axis is what characterizes the S-like configuration of the two sequences with high CG content, although the conformations are somewhat different. While poly-d(C) has a regular structure and keeps most of the canonical non-bonded interactions, poly-d(GC) presents a non-canonical 4-base stacking motif at expense of breaking some of the canonical ones.

The different types of overstretched structures obtained thanks to the use of implicitly solvated simulations were subjected to additional molecular dynamics modelling with an explicit representation of the solvent without observing any significant change. In addition, these observed structures are remarkably similar to the ones observed in previous studies [16], [17], [19], [36] confirming the improvement on the implicit description of solvent [43], [44], [60].

Our simulations show that the overstretching behaviour not only depend on AT and GC content, but also on the nature of the different type of base pair step. Thus, extrapolating these results, we think that oversimplified models base only on AT/GC ratio for describing the DNA tendency to flex beyond the canonical B-form might only be valid for a rough and ambiguous picture. For understanding molecular machines and the biomechanical function of promoter regions such as the TATA and CGCG boxes and CpG islands, it is vital to capture the behaviour of individual sequences.

**CONCLUSIONS**

The 'physics of life' is characterized by complex and emergent features with typically substantive heterogeneity of molecular components in a system. The challenges of cutting through such heterogeneity, for example as exhibited in the dynamic structural states of DNA, can be best overcome by studying one molecule at a time [61], now feasible with studies *in vitro*, in live cells and in computational simulations [62]. In this study, we have focused on a computational strategy involving molecular dynamics simulations on four distinctly different constructs of DNA in order to investigate the effect of sequence on the emergence of structural motifs in response to physiological levels of mechanical twist and stretch. We find evidence that sequence differences contribute significantly to the emergence, or not, of mechanically stable motifs. The presence of such motifs may potentially aid in recognition and transcription [63]. For each different sequence simulated even positively supercoiled DNA was found to form melting bubbles at the early stages of overstretching, indicating that, in the physiological regime of mechanical perturbation we use here, stretch is the dominant perturbation. We also find evidence that specific structural transitions occur in short ~5 bp regions, possible even in the absence of any topoisomerase activity. Our findings reinforce the importance of sequence to DNA topology, of particular relevance to sequence differences in gene promoter regions, in which mechanical perturbations have crucial roles.

**ACKNOWLEDGEMENTS**

We would like to thank Dr Phil Hasnip for many constructive discussions, the York Advanced Research Computing Cluster for computational time, and the Biological Physical Sciences Institute (BPSI), University of York for providing initial pump-priming resources.

**FUNDING**

This work was supported by funding from the Engineering and Physical Sciences Research Council (EPSRC) awards 1506874 and EP/N027639/1, Biology and Biotechnology Research Council (BBSRC) award BB/R001235/1, Leverhulme Trust award RPG-2017-340, and Royal Academy of Engineering award LTSRF1617/13/58.

**CONFLICT OF INTEREST**

There are no conflicts to declare.

**REFERENCES**

[1] A. Saha, J. Wittmeyer, and B. R. Cairns, "Chromatin remodelling: the industrial revolution of DNA around histones," *Nat. Rev. Mol. Cell Biol.*, vol. 7, p. 437, Jun. 2006.

[2] J. Méndez and B. Stillman, "Perpetuating the double helix: molecular machines at eukaryotic DNA replication origins," *BioEssays*, vol. 25, no. 12, pp. 1158–1167, 2003.

[3] S. B. Smith, Y. Cui, and C. Bustamante, "Overstretching B-DNA: The Elastic Response of Individual Double-Stranded and Single-Stranded DNA Molecules," *Science (80-. ).*, vol. 271, no. 5250, pp. 795–799, 1996.

[4] P. Cluzel *et al.*, "DNA: An Extensible Molecule," *Science (80-. ).*, vol. 271, no. 5250, p. 792 LP-794, Feb. 1996.

[5] M. Rief, H. Clausen-Schaumann, and H. E. Gaub, "Sequence-dependent mechanics of single DNA molecules," *Nat. Struct. Biol.*, vol. 6, p. 346, Apr. 1999.

[6] J. F. Leger *et al.*, "Structural transitions of a twisted and stretched DNA molecule," *Phys. Rev. Lett.*, vol. 83, no. 5, p. 1066, 1999.

[7] J. F. Allemand, D. Bensimon, R. Lavery, and V. Croquette, "Stretched and overwound DNA forms a Pauling-like structure with exposed bases," *Proc. Natl. Acad. Sci.*, vol. 95, no. 24, pp. 14152–14157, 1998.

[8] H. Miller, Z. Zhou, A. J. M. Wollman, and M. C. Leake, "Superresolution imaging of single DNA molecules using stochastic photoblinking of minor groove and intercalating dyes," *Methods*, vol. 88, pp. 81–8, Jan. 2015.

[9] H. Miller, Z. Zhou, J. Shepherd, A. J. M. Wollman, and M. C. Leake, "Single-molecule techniques in biophysics: a review of the progress in methods and applications," *Rep. Prog. Phys.*, vol. 81, no. 2, p. 24601, 2017.

[10] A. J. M. Wollman and M. C. Leake, "Millisecond single-molecule localization microscopy combined with convolution analysis and automated image segmentation to determine protein concentrations in complexly structured, functional cells, one cell at a time.," *Faraday Discuss.*, vol. 184, pp. 401–24, Jan. 2015.

[11] T. R. Strick, J.-F. Allemand, D. Bensimon, and V. Croquette, "Stress-Induced Structural Transitions in DNA and Proteins," *Annu. Rev. Biophys. Biomol. Struct.*, vol. 29, no. 1, pp. 523–543, Jun. 2000.

[12] M. W. Konrad and J. I. Bolonick, "Molecular Dynamics Simulation of DNA Stretching Is Consistent with the Tension Observed for Extension and Strand Separation and Predicts a Novel Ladder Structure," *J. Am. Chem. Soc.*, vol. 118, no. 45, pp. 10989–10994, 1996.

[13] A. Lebrun and R. Lavery, "Modelling Extreme Stretching of DNA," *Nucleic Acids Res.*, vol. 24, no. 12, pp. 2260–2267, Jun. 1996.

[14] X.-J. Lu and W. K. Olson, "Resolving the discrepancies among nucleic acid conformational analyses11Edited by I. Tinoco," *J. Mol. Biol.*, vol. 285, no. 4, pp. 1563–1575, 1999.

[15] A. D. M. Jr and G. U. Lee, "Structure, force, and energy of a double-stranded DNA oligonucleotide under tensile loads," *Eur. Biophys. J.*, vol. 28, pp. 415–426, 1999.

[16] S. Piana, "Structure and energy of a DNA dodecamer under tensile load," *Nucleic Acids Res.*, vol. 33, no. 22, pp. 7029–7038, 2005.

[17] S. A. Harris, Z. A. Sands, and C. A. Laughton, "Molecular Dynamics Simulations of Duplex Stretching Reveal the Importance of Entropy in Determining the Biomechanical Properties of DNA," *Biophys. J.*, vol. 88, no. 3, pp. 1684–1691, 2005.

[18] H. Li and T. Gisler, "Overstretching of a 30 bp DNA duplex studied with steered molecular dynamics simulation: Effects of structural defects on structure and force-extension relation," *Eur. Phys. J. E*, vol. 30, no. 3, p. 325, 2009.

[19] S. Bag, S. Mogurampelly, W. A. Goddard III, and P. K. Maiti, "Dramatic changes in DNA conductance with stretching: structural polymorphism at a critical extension," *Nanoscale*, vol. 8, no. 35, pp. 16044–16052, 2016.

[20] D. R. Roe and A. M. Chaka, "Structural Basis of Pathway-Dependent Force Profiles in Stretched DNA," *J. Phys. Chem. B*, vol. 113, no. 46, pp. 15364–15371, Nov. 2009.

[21] A. Taghavi, P. van der Schoot, and J. T. Berryman, "DNA partitions into triplets under tension in the presence of organic cations, with sequence evolutionary age predicting the stability of the triplet phase," *Q. Rev. Biophys.*, vol. 50, p. e15, 2017.

[22] J. Wereszczynski and I. Andricioaei, "On structural transitions, thermodynamic equilibrium, and the phase diagram of DNA and RNA duplexes under torque and tension," *Proc. Natl. Acad. Sci.*, vol. 103, no. 44, p. 16200 LP-16205, Oct. 2006.

[23] K. Liebl and M. Zacharias, "Unwinding Induced Melting of Double-Stranded DNA Studied by Free Energy Simulations," *J. Phys. Chem. B*, vol. 121, no. 49, pp. 11019–11030, Dec. 2017.

[24] T. Sutthibutpong *et al.*, "Long-range correlations in the mechanics of small DNA circles under topological stress revealed by multi-scale simulation," *Nucleic Acids Res.*, vol. 44, no. 19, pp. 9121–9130, 2016.

[25] G. L. Randall, L. Zechiedrich, and B. M. Pettitt, "In the absence of writhe, DNA relieves torsional stress with localized, sequence-dependent structural failure to preserve B-form," *Nucleic Acids Res.*, vol. 37, no. 16, pp. 5568–5577, 2009.

[26] T. E. Ouldridge, A. A. Louis, and J. P. K. Doye, "Structural, mechanical, and thermodynamic properties of a coarse-grained DNA model," *J. Chem. Phys*, vol. 134, 2011.

[27] T. A. Knotts, N. Rathore, D. C. Schwartz, and J. J. de Pablo, "A coarse grain model for DNA," *J. Chem. Phys.*, vol. 126, no. 8, p. 84901, Feb. 2007.

[28] F. Romano, D. Chakraborty, J. P. K. Doye, T. E. Ouldridge, and A. A. Louis, "Coarse-grained simulations of DNA overstretching," *J. Chem. Phys.*, vol. 138, no. 8, p. 85101, 2013.

[29] Y.-L. Zhu, Z.-Y. Lu, and Z.-Y. Sun, "The mechanism of the emergence of distinct

overstretched DNA states," *J. Chem. Phys.*, vol. 144, no. 2, p. 24901, 2016.

[30] A. Bancaud *et al.*, "Structural plasticity of single chromatin fibers revealed by torsional manipulation," *Nat. Struct. &Amp; Mol. Biol.*, vol. 13, p. 444, Apr. 2006.

[31] E. L. Zechiedrich *et al.*, "Roles of Topoisomerases in Maintaining Steady-state DNA Supercoiling in Escherichia coli ," *J. Biol. Chem.* , vol. 275, no. 11, pp. 8103–8113, Mar. 2000.

[32] A. Sarai, J. Mazur, R. Nussinov, and R. L. Jernigan, "Sequence dependence of DNA conformational flexibility," *Biochemistry*, vol. 28, no. 19, pp. 7842–7849, Sep. 1989.

[33] P. J. Mitchell and R. Tjian, "Transcriptional regulation in mammalian cells by sequence-specific DNA binding proteins," *Science (80-. ).*, vol. 245, no. 4916, pp. 371–378, 1989.

[34] T. Yang and B. W. Poovaiah, "A calmodulin-binding/CGCG box DNA-binding protein family involved in multiple signaling pathways in plants," *J. Biol. Chem.*, vol. 277, no. 47, pp. 45049–45058, 2002.

[35] D. Takai and P. A. Jones, "Comprehensive analysis of CpG islands in human chromosomes 21 and 22," *Proc. Natl. Acad. Sci.*, vol. 99, no. 6, pp. 3740–3745, 2002.

[36] A. Lebrun, Z. Shakked, and R. Lavery, "Local DNA stretching mimics the distortion caused by the TATA box-binding protein," *Proc. Natl. Acad. Sci.*, vol. 94, no. 7, p. 2993 LP-2998, Apr. 1997.

[37] Z. Chen, H. Yang, and N. P. Pavletich, "Mechanism of homologous recombination from the RecA–ssDNA/dsDNA structures," *Nature*, vol. 453, p. 489, May 2008.

[38] D. A. Case *et al.*, "AMBER17. 2017," *San Fr. Univ. Calif.*, 2018.

[39] T. E. Cheatham, P. Cieplak, and P. A. Kollman, "A Modified Version of the Cornell et al. Force Field with Improved Sugar Pucker Phases and Helical Repeat," *J. Biomol. Struct. Dyn.*, vol. 16, no. 4, pp. 845–862, Feb. 1999.

[40] A. Pérez *et al.*, "Refinement of the AMBER force field for nucleic acids: improving the description of alpha/gamma conformers," *Biophys. J.*, vol. 92, no. 11, pp. 3817–3829, 2007.

[41] I. Ivani *et al.*, "Parmbsc1: a refined force field for DNA simulations," *Nat. Methods*, vol. 13, no. 1, p. 55, 2016.

[42] J. Srinivasan, M. W. Trevathan, P. Beroza, and D. A. Case, "Application of a pairwise generalized Born model to proteins and nucleic acids: inclusion of salt effects," *Theor. Chem. Acc.*, vol. 101, no. 6, pp. 426–434, 1999.

[43] H. Nguyen, D. R. Roe, and C. Simmerling, "Improved Generalized Born Solvent Model Parameters for Protein Simulations," *J. Chem. Theory Comput.*, vol. 9, no. 4, pp. 2020–2034, Apr. 2013.

[44] H. Nguyen, A. Peerez, S. Bermeo, and C. Simmerling, "Refinement of generalized born implicit solvation parameters for nucleic acids and their complexes with proteins," *J. Chem. Theory Comput.*, vol. 11, no. 8, pp. 3714–3728, 2015.

[45] R. J. Loncharich, B. R. Brooks, and R. W. Pastor, "Langevin dynamics of peptides: The

frictional dependence of isomerization rates of n-acetylalnyl-n'-methylamide," *Biopolymers*, vol. 32, pp. 523–535, 1992.

[46] A. Noy and R. Golestanian, "Length Scale Dependence of DNA Mechanical Properties," *Phys. Rev. Lett.*, vol. 109, no. 22, p. 228101, Nov. 2012.

[47] D. R. Roe and T. E. Cheatham III, "PTRAJ and CPPTRAJ: software for processing and analysis of molecular dynamics trajectory data," *J. Chem. Theory Comput.*, vol. 9, no. 7, pp. 3084–3095, 2013.

[48] W. Humphrey, A. Dalke, and K. Schulten, "{VMD} -- {V}isual {M}olecular {D}ynamics," *J. Mol. Graph.*, vol. 14, pp. 33–38, 1996.

[49] E. F. Pettersen *et al.*, "UCSF Chimera—a visualization system for exploratory research and analysis," *J. Comput. Chem.*, vol. 25, no. 13, pp. 1605–1612, 2004.

[50] D. J. Price and C. L. Brooks, "A modified TIP3P water potential for simulation with Ewald summation," *J. Chem. Phys.*, vol. 121, no. 20, pp. 10096–10103, Nov. 2004.

[51] I. S. Joung and T. E. Cheatham, "Determination of Alkali and Halide Monovalent Ion Parameters for Use in Explicitly Solvated Biomolecular Simulations," *J. Phys. Chem. B*, vol. 112, no. 30, pp. 9020–9041, Jul. 2008.

[52] I. S. Joung and T. E. Cheatham, "Molecular Dynamics Simulations of the Dynamic and Energetic Properties of Alkali and Halide Ions Using Water-Model-Specific Ion Parameters," *J. Phys. Chem. B*, vol. 113, no. 40, pp. 13279–13290, Oct. 2009.

[53] A. Noy, A. Maxwell, and S. A. Harris, "Interference between Triplex and Protein Binding to Distal Sites on Supercoiled DNA," *Biophys. J.*, vol. 112, no. 3, pp. 523–531, 2017.

[54] H. J. C. Berendsen, J. P. M. Postma, W. F. van Gunsteren, A. DiNola, and J. R. Haak, "Molecular dynamics with coupling to an external bath," *J. Chem. Phys.*, vol. 81, no. 8, pp. 3684–3690, Oct. 1984.

[55] F. Kilchherr, C. Wachauf, B. Pelz, M. Rief, M. Zacharias, and H. Dietz, "Single-molecule dissection of stacking forces in DNA," *Science (80-. ).*, vol. 353, no. 6304, p. aaf5508, Sep. 2016.

[56] J. Šponer, P. Jurečka, I. Marchan, F. J. Luque, M. Orozco, and P. Hobza, "Nature of Base Stacking: Reference Quantum-Chemical Stacking Energies in Ten Unique B-DNA Base-Pair Steps," *Chem. – A Eur. J.*, vol. 12, no. 10, pp. 2854–2865, Mar. 2006.

[57] R. Vlijm, J. v.d. Torre, and C. Dekker, "Counterintuitive DNA Sequence Dependence in Supercoiling-Induced DNA Melting," *PLoS One*, vol. 10, no. 10, p. e0141576, Oct. 2015.

[58] T. Strick, J.-F. Allemand, D. Bensimon, R. Lavery, and V. Croquette, "Phase coexistence in a single DNA molecule," *Phys. A Stat. Mech. its Appl.*, vol. 263, no. 1, pp. 392–404, 1999.

[59] R. Lohikoski, J. Timonen, and A. Laaksonen, "Molecular dynamics simulation of single DNA stretching reveals a novel structure," *Chem. Phys. Lett.*, vol. 407, no. 1, pp. 23–29, 2005.

[60] A. Perez, J. L. MacCallum, E. Brini, C. Simmerling, and K. A. Dill, "Grid-Based Backbone Correction to the ff12SB Protein Force Field for Implicit-Solvent Simulations," *J. Chem.*

*Theory Comput.*, vol. 11, no. 10, pp. 4770–4779, Oct. 2015.

[61] M. C. Leake, "The physics of life: one molecule at a time.," *Philos. Trans. R. Soc. Lond. B. Biol. Sci.*, vol. 368, no. 1611, p. 20120248, Feb. 2013.

[62] A. J. M. Wollman, H. Miller, Z. Zhou, and M. C. Leake, "Probing DNA interactions with proteins using a single-molecule toolbox: inside the cell, in a test tube and in a computer," *Biochem. Soc. Trans.*, vol. 43, no. 2, pp. 139–145, 2015.

[63] N. A. Davis, S. S. Majee, and J. D. Kahn, "TATA Box DNA Deformation with and without the TATA Box-binding Protein," *J. Mol. Biol.*, vol. 291, no. 2, pp. 249–265, 1999.

**Supplementary Information**

**The emergence of sequence-dependent structural motifs in stretched, torsionally constrained DNA**

Jack W Shepherd[1, †], R J Greenall[1], M I J Probert[1], Agnes Noy[1*], Mark C. Leake[1,2*]

[1] Department of Physics, University of York, York, YO10 5DD, UK.

[2] Department of Biology, University of York, York, YO10 5NG, UK.

* To whom correspondence should be addressed, Email: mark.leake@york.ac.uk
Tel: +44 (0)1904 322697; Fax: +44 (0)1904 322214.
Correspondence may also be addressed to Agnes Noy, Email: agnes.noy@york.ac.uk

[†] Present Address: Jack W Shepherd, Cell Polarity Migration and Cancer Unit, Institut Pasteur, 25-28 Rue de Dr Roux, 75015 Paris, France.

**Supplementary Figures**

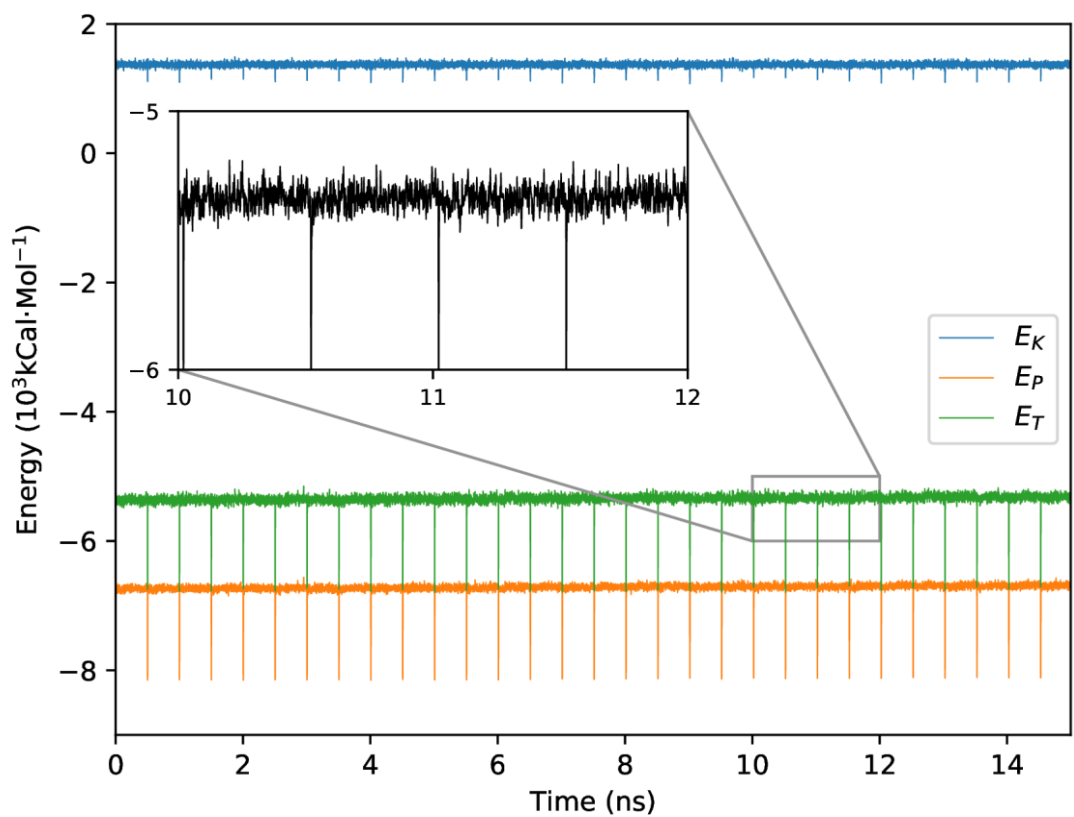

**Figure S1:** Total, potential, and kinetic energies over a full stretching simulation of of poly-d(A)·poly-d(T) with σ=- 0.064 stretched by 3 nm over 15 ns, with a sudden stretch of 0.1 nm every 500 ps. The energies rapidly return to an equilibrium level after the stretching event, demonstrating that the energy minimisation procedure has equilibrated the system effectively.

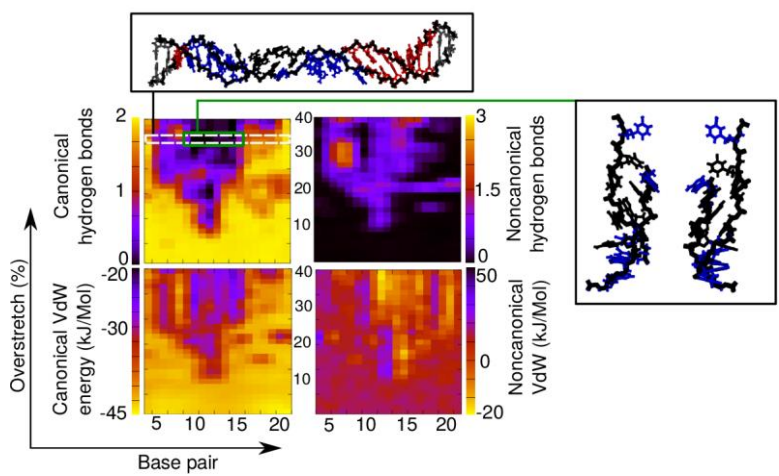

**Figure S2**: Stacking and hydrogen bonding for (AA)$_{12}$ with σ=0.068.

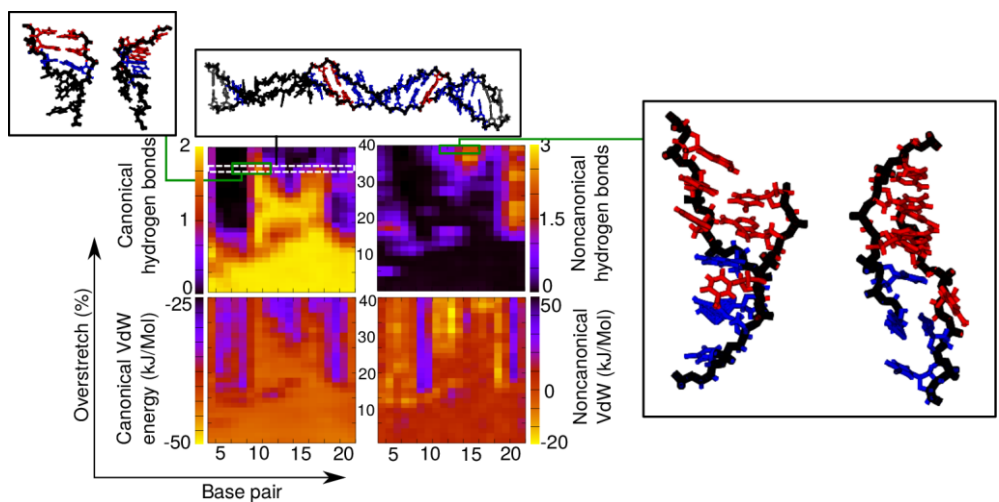

**Figure S3**: Stacking, hydrogen bonding for (AT)$_{12}$ with σ=0.068.

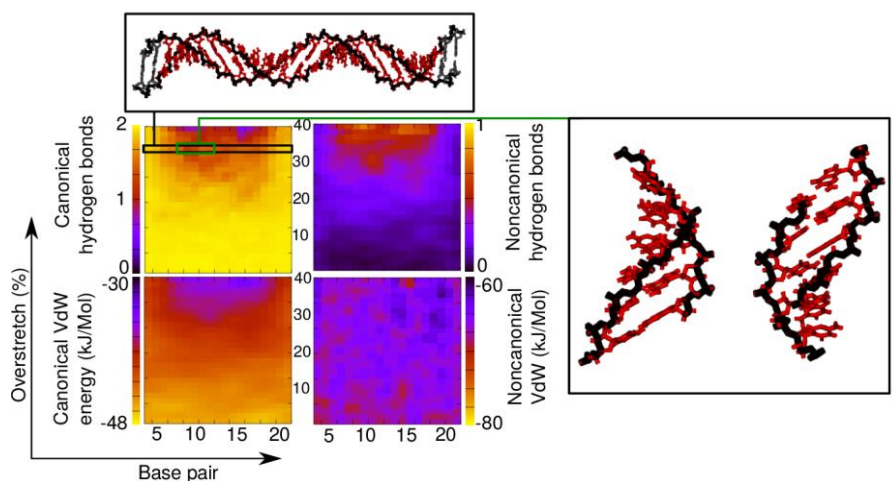

**Figure S4**: Stacking, hydrogen bonding, and motifs for (CC)$_{12}$ overtwisted to σ=0.068.

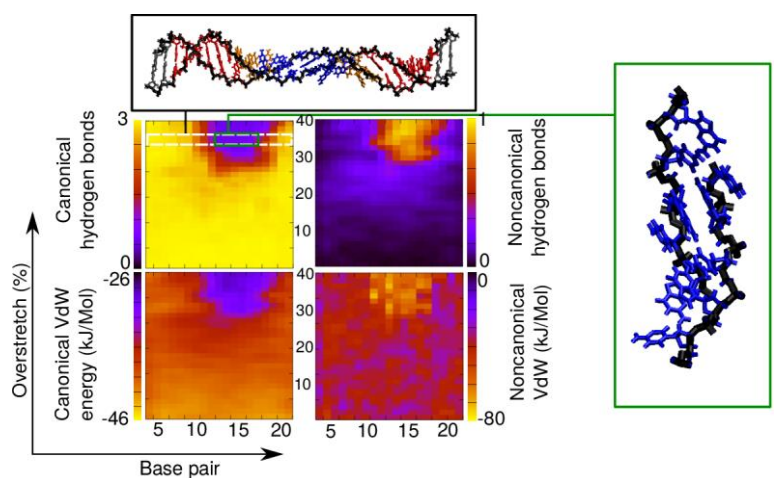

**Figure S5**: Stacking, hydrogen bonding, and motifs for (CG)$_{12}$ where σ=0.068.

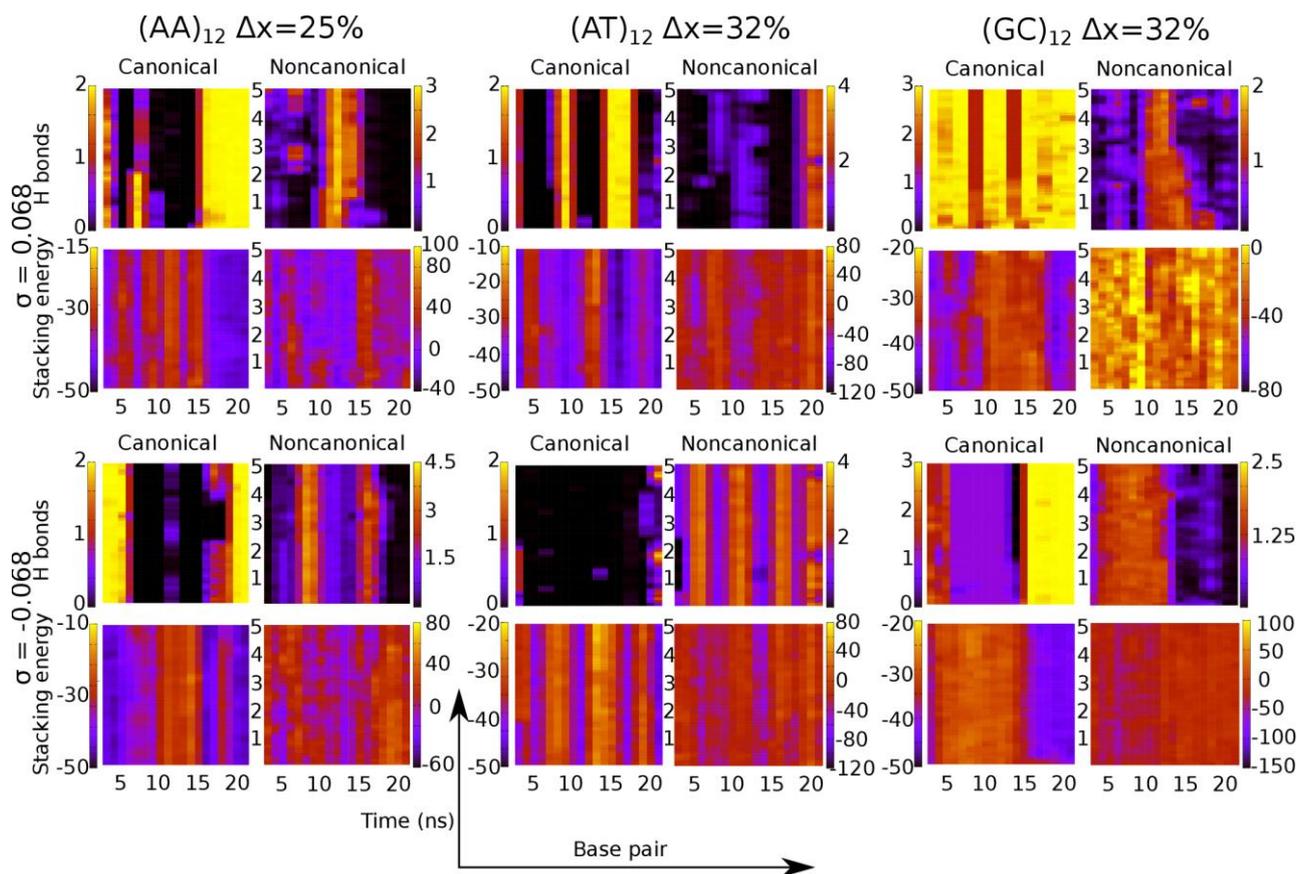

**Figure S6**: Hydrogen bonding and stacking interactions through 5 ns of explicit solvent simulation in 200 mM NaCl for six structures. Leftmost panels: (AA)$_{12}$ with a relative overextension of 25% shows maintains the denaturation bubble and noncanonical hydrogen bonding/stacking. Centre panels: (AT)$_{12}$ retains the distinctive pattern of hydrogen bonds and stacking interactions with separate regions engaged in either more noncanonical stacking or more noncanonical hydrogen bonding. This behaviour is conserved in both the over and undertwisted case. Right panels: (GC)$_{12}$ partitions perturbations efficiently for both supercoiling densities, leading to a coexistence of conformations within the molecule. Structures taken from these simulations can be seen in Figure S7

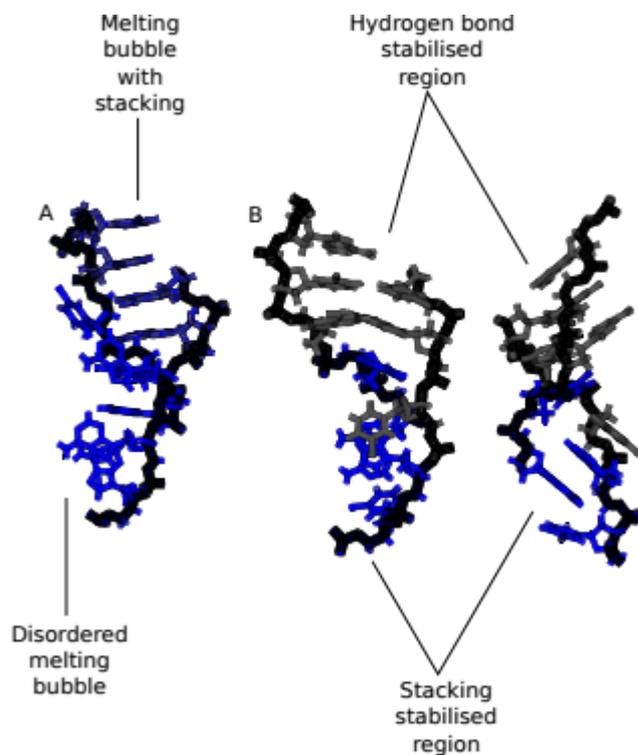

**Figure S7:** A) Melting bubbles seen after averaging 5 ns of explicit solvent simulation of poly-d(A).poly-d(T) with σ=0.068 and relative overextension 28%. Two melting bubble modes are in evidence, one where there is non-canonical stacking as in Figure 2, and one which is more disordered. B) The coexisting stabilisation modes of poly-d(AT).poly-d(AT) with σ=-0.068 and relative overextension 35%, averaged over 5 ns of explicit solvent simulation. The non-canonical hydrogen bonding and non-canonical stacking persist over the course of the simulation. Both simulations were performed with 200 mM NaCl.